\documentclass[11pt,a4paper]{article}

\usepackage{amsmath,amssymb,amsfonts,graphics,graphicx,amscd,amsfonts,epsfig,color}
\usepackage{jheppub} 
\usepackage{amssymb,amsmath}
\graphicspath{{./Fig/}}

\usepackage{graphicx}
\usepackage{color}

\newcommand{\SUN}[1]{{\rm SU}(N)}

\newcommand {\beq} {\begin{equation}}
\newcommand {\eeq} {\end{equation}}
\newcommand {\beqa}{\begin{eqnarray}}
\newcommand {\eeqa}{\end{eqnarray}}

\newcommand {\tr}{{\rm tr\,}}

\newcommand{\bbC}{{\mathbb C}}

\begin{document}

\title{Testing the criterion for correct convergence
in the complex Langevin method
}

\author[a,b]{Keitaro Nagata,}
\author[b,c]{Jun Nishimura}
\author[b,d]{and Shinji Shimasaki}

\affiliation[a]{Center of Medical Information Science, 
Kochi Medical School, Kochi University,\\
Kohasu, Oko-cho, Nankoku-shi, Kochi 783-8505, Japan}
\affiliation[b]{KEK Theory Center,
High Energy Accelerator Research Organization,\\
1-1 Oho, Tsukuba, Ibaraki 305-0801, Japan}
\affiliation[c]{Department of Particle and Nuclear Physics, 
School of High Energy Accelerator Science,\\
Graduate University for Advanced Studies (SOKENDAI),\\
1-1 Oho, Tsukuba, Ibaraki 305-0801, Japan}
\affiliation[d]{Research and Education Center for Natural Sciences, 
Keio University,\\
Hiyoshi 4-1-1, Yokohama, Kanagawa 223-8521, Japan
}

\emailAdd{k-nagata@kochi-u.ac.jp}
\emailAdd{jnishi@post.kek.jp}
\emailAdd{shinji.shimasaki@keio.jp}
\note{KEK-TH-2032}

\date{\today
}

\abstract{Recently the complex Langevin method (CLM) has been attracting
attention as a solution to the sign problem, which occurs in 
Monte Carlo calculations when the effective Boltzmann weight is not
real positive. 
An undesirable feature of the method, however, was that it 
can happen in some parameter regions
that the method yields wrong results even if the Langevin process 
reaches equilibrium
without any problem.
In our previous work, we proposed a practical criterion 
for correct convergence based on the probability distribution of the
drift term that appears in the complex Langevin equation.
Here we demonstrate the usefulness of this criterion in two
solvable theories with many dynamical degrees of freedom, i.e., 
two-dimensional Yang-Mills theory with a complex coupling constant
and the chiral Random Matrix Theory 
for finite density QCD,
which were studied by the CLM before.
Our criterion can indeed tell the parameter regions in which 
the CLM gives correct results.
}

\maketitle

\section{Introduction}

The sign problem is one of the most important issues in 
contemporary physics, 
which hinders theoretical developments
in QCD at finite density,
real-time dynamics of quantum many-body systems,
strongly coupled electron systems,
supersymmetric theories and so on.
In the path-integral formulation, these theories
typically have an effective Boltzmann weight which is not real positive,
and hence the importance sampling
used in conventional Monte Carlo methods does not work.
The complex Langevin method (CLM) is a promising candidate 
of the methods that can be 
applied in such cases.
It is based on the stochastic 
quantization \cite{Parisi:1980ys,Damgaard:1987rr},
which uses a Langevin process 
associated with the Boltzmann weight.
Since it
does not rely on the probabilistic interpretation 
of the Boltzmann weight,
it has a chance to be
generalized to the case of a complex Boltzmann 
weight~\cite{Parisi:1984cs,Klauder:1983sp},
which, however, necessarily requires 
the dynamical variables 
that are real in the original theory to be complexified.
Accordingly, the observables and 
the drift term in the Langevin process 
are defined
for complexified variables
by analytic continuation.
%

While the Langevin method as applied to 
a system with a real positive Boltzmann weight
yields correct results in general,
it is known that the CLM
does not always yield correct results,
and this feature had not been understood for quite a long time.
An important progress was made by 
refs.~\cite{Aarts:2009uq, Aarts:2011ax},
in which the justification of the CLM was discussed
based on an equality between the expectation value of 
observables defined in the CLM and the expectation value 
defined in the original path integral.
It was noticed that the integration by parts used to prove the
equality cannot be justified if
the complexified variables make long excursions 
in the imaginary directions (``the excursion problem'').
The same obstacle appears when
the drift term has singularities and 
the complexified variables come close to 
these singularities frequently \cite{Nishimura:2015pba} 
(``the singular drift problem'').
Thus the reasons for wrong convergence in the CLM 
was understood at least theoretically.
%
For recent progress concerning the CLM,
see refs.~\cite{Seiler:2012wz,Sexty:2013ica,Fukushima:2015qza,Nagata:2015uga,Tsutsui:2015tua,Fodor:2015doa,Hayata:2015lzj,Ichihara:2016uld,Aarts:2016qrv,Nagata:2016vkn,Abe:2016hpd,Aarts:2016qhx,Ito:2016efb,Bloch:2017ods,Aarts:2017vrv,Seiler:2017wvd,Doi:2017gmk,Nagata:2017pgc,Sinclair:2017zhn,Fujii:2017oti,Bloch:2017sex,Anagnostopoulos:2017gos}.

A crucial issue in the CLM is therefore 
how one can judge 
whether these problems are occurring or not during the simulation.
In ref.~\cite{Aarts:2009uq}, the Langevin-time evolution operator 
$\tilde{L}$ acting on an observable ${\cal O}$ was considered,
and the identity $\langle \tilde{L}{\cal O} \rangle = 0$
in the long-time limit
was proposed as a necessary condition for the validity of
integration by parts used in the justification.
While this criterion was shown to be useful in simple models, 
it is numerically demanding to apply it to models with many
dynamical degrees of freedom
since the quantity $\tilde{L}{\cal O}$ fluctuates
violently around zero, and it requires a tremendous amount of 
statistics in order to judge whether it averages to zero or not.
One should also note that the integration by part is not
fully justified even if this criterion is met
because it is merely a necessary condition.

Recently, we have reconsidered the argument for justification
of the CLM \cite{Nagata:2016vkn}, and
pointed out a subtlety in the use of time-evolved observables,
which plays a crucial role in the argument.
Our refined argument, which cures this subtlety,
requires
the probability distribution of the drift term
to fall off exponentially or faster at large magnitude.
The issue of the integration by parts can actually be reformulated
in terms of the same probability distribution, and the corresponding
condition turned out to be slightly weaker than the one above.
Thus the above condition was proposed as
a necessary and sufficient condition 
for correct convergence in the CLM under obvious assumptions
such as the stability
of the Langevin process\footnote{Based on studies of simple models,
it has been emphasized recently that the ergodicity of the 
Langevin process is also an important 
assumption \cite{Aarts:2017vrv,Seiler:2017wvd}.}
and the convergence of the
observable itself.
Since the drift term is a quantity that one has to calculate anyway
at each Langevin step, probing its distribution 
costs almost nothing in addition.
In the same paper, we have shown
the validity of our criterion
in simple one-variable models.

Our criterion may be viewed as a refinement of 
the theoretical understanding
that the probability distribution of the dynamical variables 
should decay fast enough at infinity \cite{Aarts:2009uq, Aarts:2011ax}
and at the singularities of 
the drift term \cite{Nishimura:2015pba,Aarts:2017vrv}.
However, the statement based on the magnitude of the drift term
has a big advantage that it enables us to claim how fast the distribution
should decay for the correct convergence.

The purpose of this work is to demonstrate the usefulness
of our criterion in models with many dynamical degrees of freedom.
Here we study two solvable models,
two-dimensional pure Yang-Mills 
theory (2dYM) \cite{Balian:1974ts,Migdal:1975zg,Rothe:1992nt} 
with a complex coupling constant
and
the chiral Random Matrix Theory (cRMT) 
for finite density QCD \cite{Osborn:2004rf,Bloch:2012bh},
which were studied by the CLM
in refs.~\cite{Makino:2015ooa}
and 
\cite{Mollgaard:2013qra,Mollgaard:2014mga,Nagata:2015ijn,Nagata:2016alq}, 
respectively.
In both models, the CLM reproduced the exact results correctly
in some parameter region but not in the other,
due to the excursion problem and the singular-drift problem, respectively.
Since the results of the CLM depend smoothly on the parameter,
it was not possible to identify precisely 
the parameter region in which the CLM is valid 
without knowing the exact results.
Our results 
for the probability distribution of the drift term 
indeed show a drastic change of its behavior at large magnitude 
as expected depending on the parameter regions.
This demonstrates the usefulness of our criterion 
for correct convergence in the CLM.

The rest of this paper is organized as follows. 
In section \ref{sec:correct convergence},
we briefly review the CLM.
In particular, we discuss the criterion
for correct convergence proposed in ref.~\cite{Nagata:2016vkn}
and the gauge cooling technique used in the present work.
In sections \ref{sec:2dYM} and \ref{sec:cRMT}, 
we apply the CLM to the 2dYM with a complex coupling
constant and the cRMT for finite density QCD, respectively.
In particular, we provide numerical results which demonstrate
the usefulness of our criterion.
Section \ref{sec:summary} is devoted to a summary and discussions.

\section{Brief review of the complex Langevin method}
\label{sec:correct convergence}

In this section, we review the CLM and the criterion for correct convergence
using a system
\begin{align}
  Z=\int dx \, w(x) 
\label{original-theory}
\end{align}
of $N$ real variables $x_k$ ($k=1 ,\cdots , N$)
as a simple example. 
Here the weight $w(x)$ is a complex-valued function, 
which causes the sign problem.

\subsection{complex Langevin method}


In the CLM, 
the original real variables $x_k$ 
are complexified as $x_k\to z_k=x_k+iy_k\in \mathbb{C}$
and one considers a fictitious time evolution of
the complexified variables $z_k$ using
the complex Langevin equation
given, in its discretized form, by
\begin{align}
z^{(\eta)}_k(t+\epsilon)=z^{(\eta)}_k(t)
+\epsilon \, v_k(z^{(\eta)}(t))+\sqrt{\epsilon} \, \eta_k(t) \ ,
\label{Langevin}
\end{align}
where $t$ is the fictitious time with a stepsize $\epsilon$.
The second term $v_k(z)$ on the right-hand side is called 
the drift term, which is defined by
holomorphic extension of the one
\begin{align}
v_k(x)=w(x)^{-1}\frac{\partial w(x)}{\partial x_k}
\end{align}
for the real variables $x_k$.
The variables $\eta_k(t)$ appearing on the right-hand side of 
eq.~(\ref{Langevin})
are a real Gaussian noise 
with the probability distribution
$\propto e^{-\frac{1}{4}\sum_t\eta_k(t)^2}$,
which makes the time-evolved variables $z^{(\eta)}_k(t)$ stochastic.
The expectation values
with respect to the noise $\eta_k(t)$ 
are denoted as $\langle \cdots \rangle_\eta$ in what follows.

Let us consider the expectation value of 
an observable $\mathcal O(x)$.
In the CLM, one computes the expectation value of the 
holomorphically extended observable $\mathcal O(x+iy)$ as
\begin{align}
  \Phi(t)= \Big\langle \mathcal O(z^{(\eta)}(t)) \Big\rangle_\eta
=\int dx \, dy \, \mathcal O(x+iy) P(x,y;t) \ ,
  \label{Phi}
\end{align}
where $P(x,y;t)$ is the probability distribution of 
$x^{(\eta)}(t)$ and $y^{(\eta)}(t)$ defined by
\begin{align}
  P(x,y;t)=\Big \langle 
\delta(x-x^{(\eta)}(t))\delta(y-y^{(\eta)}(t)) \Big\rangle_\eta \ .
  \label{P}
\end{align}
Then, the correct convergence of the CLM implies the equality
\begin{align}
  \lim_{t\to \infty}\lim_{\epsilon \to 0}\Phi(t)
=\frac{1}{Z}\int dx \, \mathcal O (x)w(x) \ ,
  \label{key}
\end{align}
where the right-hand side is the expectation value 
of $\mathcal O(x)$ in the original theory (\ref{original-theory}).
A proof of eq.~\eqref{key} was given
in refs.~\cite{Aarts:2009uq,Aarts:2011ax}, where the notion of
the time-evolved observable $\mathcal O (z;t)$ plays a crucial role.
In particular, it was pointed out that the integration by parts 
used in the argument cannot be justified 
when the probability distribution \eqref{P} falls off slowly
in the imaginary direction.
In ref.~\cite{Nishimura:2015pba},
it was noticed that the wrong convergence
associated with the zeroes of the fermion determinant
\cite{Mollgaard:2013qra}
is actually due to 
the slow fall-off of the probability distribution \eqref{P} 
toward the singularities of the drift term.

While this argument provided theoretical understanding of the cases
in which the CLM gives wrong results, the precise condition on the 
probability distribution was not specified.
Furthermore, there is actually a subtlety
in defining the time-evolved observable $\mathcal O (z;t)$ 
as we discuss in the next subsection.

\subsection{the condition for correct convergence}
\label{Condition for the correct convergence}

Here we review the refined argument for justification
of the CLM, which leads to the condition for 
correct convergence \cite{Nagata:2016vkn}.

The basic idea in proving the equality \eqref{key}
is to consider the time evolution of the expectation value $\Phi(t)$, 
which is given by
\begin{align}
\Phi(t + \epsilon)  &= 
\int 
dx \, dy \, {\cal O}_{\epsilon}(x+iy) \, 
P(x,y;t) \ ,
\label{OP-rewriting-P2}
\end{align}
where we have defined the time-evolved observable
\begin{align}
{\cal O}_{\epsilon}(z) 
&=   \frac{1}{ {\cal N}}
\int d\eta \, 
e^{-\frac{1}{4}  \eta^{2} } 
{\cal O} \Big(z+\epsilon \, v(z)+\sqrt{\epsilon}\, \eta  \Big) \ .
\label{OP-rewriting-P3}
\end{align}
Note that if $\mathcal{O}(z)$ and $v(z)$ are holomorphic, 
so is $\mathcal O_{\epsilon}(z)$.
Expanding the right-hand side of (\ref{OP-rewriting-P3})
with respect to $\epsilon$
and integrating $\eta$ out, one can rewrite 
(\ref{OP-rewriting-P2}) as
\begin{align}
  \Phi(t+\epsilon)
  &= 
\sum_{n=0}^{\infty}
  \frac{1}{n!} \, \epsilon^n  
\int 
dx \,  dy \, 
\Big\{ \mbox{\bf :} \tilde{L}^n  \mbox{\bf :} \,  {\cal O}(z) \Big\}
P(x,y;t) \ ,
\label{OP-rewriting-P3b}
\end{align}
where we have defined a differential operator
\begin{align}
  \tilde{L}
  = \left( \frac{\partial}{\partial z_k} + v_k(z)
  \right)
  \frac{\partial}{\partial z_k}
\label{def-tildeL}
\end{align}
acting on a holomorphic function of $z_k$,
and the symbol $\mbox{\bf :}\cdots\mbox{\bf :}$ 
implies that the derivatives 
are moved to the right, i.e., $\mbox{\bf :} ( f(x) + \partial)^2 \mbox{\bf :} 
= f(x)^2 + 2f(x)\partial + \partial^2$.

Taking the $\epsilon\to 0$ limit in (\ref{OP-rewriting-P3b}), 
one naively obtains
\begin{align}
  \frac{d}{dt} \, \Phi(t)
&=  \int dx \, dy \, 
\Big\{ \tilde{L} \, {\cal O}(z) \Big\}
\, P(x,y;t)  
\label{OP-rewriting-P3b-cont-lim}
\end{align}
and a finite time evolution of $\Phi(t)$ as
\begin{align}
 \Phi(t+\tau)
&=  \sum_{n=0}^{\infty}
  \frac{1}{n!} \, \tau^n
\int dx \, dy \, 
\Big\{  \tilde{L}^n \, {\cal O}(z) \Big\}
\, P(x,y;t)  \ .
\label{OP-rewriting-P3b-cont-lim-exp}
\end{align}
Assuming that \eqref{OP-rewriting-P3b-cont-lim-exp} is valid 
for finite $\tau$ at arbitrary $t$,
one can derive
the time evolution of 
an equivalent system of real variables by induction
with respect to $t$, from which eq.~\eqref{key} follows.

The expressions such as
\eqref{OP-rewriting-P3b} and \eqref{OP-rewriting-P3b-cont-lim-exp}
need some care, though.
In order for the $\epsilon$-expansion
\eqref{OP-rewriting-P3b} to be valid,
the integral on the right-hand side should be convergent for all $n$.
In order for the expression \eqref{OP-rewriting-P3b-cont-lim-exp} to be 
valid for finite $\tau$,
the integral on the right-hand side should be convergent for all $n$,
and on top of that,
the infinite sum over $n$ should have a finite convergence radius,
which may depend on $t$.

The issues raised above are nontrivial since the drift term $v_k(z)$
in the differential operator (\ref{def-tildeL}) can become large
for some $z=x+iy$, which appears with the probability distribution 
$P(x,y;t)$. Defining the magnitude of the drift term $u(z)$ in 
a suitable manner, the most dominant contribution
from $\tilde{L}^n$ in 
\eqref{OP-rewriting-P3b} and \eqref{OP-rewriting-P3b-cont-lim-exp}
can be estimated as $\tilde{L}^n \sim u(z)^n$.
Therefore, the integral appearing in the infinite series
can be estimated as
\begin{align}
  \int dx \, dy \, u(z)^n \, P(x,y;t) 
= \int_0^\infty  du \, u^n \, p(u;t) \ ,
\label{simplified-integral}
\end{align}
where we have defined the probability distribution of $u(z)$ by
\begin{align}
  p(u;t) \equiv  \int dx \, dy \, \delta(u(z)-u) \, P(x,y;t) \ .
  \label{def-u-prob}
\end{align}
In order for \eqref{OP-rewriting-P3b-cont-lim} to be valid, 
\eqref{simplified-integral} should be finite for arbitrary $n$,
which requires that $p(u;t)$ should fall off faster than any power law.
In order for the infinite series \eqref{OP-rewriting-P3b-cont-lim-exp}
to have a finite convergence radius,
$p(u;t)$ should fall off exponentially or faster.
Since the latter condition is slightly stronger than the former,
it can be regarded as 
a necessary and sufficient condition for correct convergence in the CLM.

In the previous argument \cite{Aarts:2009uq,Aarts:2011ax},
eq.~\eqref{OP-rewriting-P3b-cont-lim} 
was derived in a continuous time formulation,
where the time evolution of the probability distribution $P(x,y;t)$ 
was converted to the time evolution of the observable
using integration by parts.
If the integration by parts can be justified and 
eq.~\eqref{OP-rewriting-P3b-cont-lim} indeed holds,
the right-hand side of \eqref{OP-rewriting-P3b-cont-lim} should vanish
in the long time limit.
This was proposed as a necessary condition for 
correct convergence in the CLM.
On the other hand, it was implicitly assumed that
the infinite series in \eqref{OP-rewriting-P3b-cont-lim-exp} has
an infinite convergence radius.
This assumption is actually too strong, though,
as we have discussed above.
Our refined argument based on induction only requires that
the infinite series in \eqref{OP-rewriting-P3b-cont-lim-exp} should have
a finite convergence radius at arbitrary $t$.
This leads to a condition, which is slightly stronger than
the condition required for justifying the integration by parts
as one can see from our derivation of \eqref{OP-rewriting-P3b-cont-lim} 
based on the $\epsilon$-expansion.

As we mentioned in the previous subsection,
the situation in which the CLM fails can be classified into two cases.
One is the case in which the complexified variables make long excursions
in the imaginary directions 
(the excursion problem) \cite{Aarts:2009uq,Aarts:2011ax},
and the other is the case in which 
the drift term has singularities and the complexified variables come 
close to these points frequently 
(the singular-drift problem) \cite{Nishimura:2015pba}.
In both these cases,
the magnitude of the drift term tends to become large, 
and the probability distribution of the drift term can have 
a power-law behavior at large magnitude.
Thus, our criterion can detect these two problems in a unified manner,
and more importantly, it enables us to determine precisely
the parameter region
in which 
these problems occur.
The usefulness of our criterion was demonstrated in 
ref.~\cite{Nagata:2016vkn}
for two simple one-variable models, which suffer from 
the excursion problem and the singular drift problem,
respectively, in some parameter region.
Whether it is useful
also for systems with many degrees of freedom
is the issue we address in what follows.

\subsection{gauge cooling}

In the present work, we use the so-called gauge cooling
to reduce the excursion problem or the singular-drift problem
as much as possible.
Here we briefly review the basic idea of this technique
using the system (\ref{original-theory}) as a simple example.

Suppose the original system (\ref{original-theory})
has a symmetry under $x_k'=g_{kl}x_l$, 
where $g_{kl}$ is an element of a Lie group.
Upon complexification $x_k\to z_k$,  
the symmetry of the action and the observables is
enhanced to $z_k' = g_{kl} \, z_l$,
where $g_{kl}$ is an element of a Lie group obtained by 
complexifying the original Lie group.
Using this fact, one can improve the Langevin process (\ref{Langevin}) as
\begin{align}
\tilde z^{(\eta)}_k(t)&=g_{kl} \, z^{(\eta)}_l(t) \ , \\
z^{(\eta)}_k(t+\epsilon)&=\tilde z_k^{(\eta)}(t)
+\epsilon \, v_k(\tilde z^{(\eta)}(t))+\sqrt{\epsilon} \, \eta_k(t) \ ,
\end{align}
where the first line represents the gauge cooling.
At each Langevin step, one chooses an appropriate transformation 
function $g$ depending on the previous configuration
in such a way that possible problems of the CLM
are avoided. 

This gauge cooling was originally proposed as a technique to 
solve the excursion problem \cite{Seiler:2012wz},
but later it was shown to be useful also in solving 
the singular-drift problem \cite{Nagata:2015ijn,Nagata:2016alq}.
Theoretical justification of this technique was given 
explicitly in refs.~\cite{Nagata:2015uga,Nagata:2016vkn}.

\section{2d Yang-Mills theory with a complex coupling constant}
\label{sec:2dYM}

In this section, we apply the CLM to 2dYM 
with a complex coupling constant,
which suffers from the excursion problem 
in some parameter region \cite{Makino:2015ooa}.



\subsection{the model}

Let us consider 2dYM with an SU($N_{\rm c}$) gauge group,
which is defined by
\begin{align}
Z&=\int {\cal D} U e^{-S(U)}, \\
S&=-\frac{\beta}{2N_{\rm c}} \sum_n {\rm tr} 
\biggl[ U_{12}(n) + U_{12}^\dagger(n)\biggr] \ ,
\label{eq:action_orig}
\end{align}
where $n$ represents a site on a 
$N_{\rm t}\times N_{\rm s}$ lattice with periodic boundary conditions 
and $U_{12}(n)$ represents the plaquette defined in terms of the 
link variables $U_{n\mu}\in \text{SU}(N_{\rm c})$
by $U_{12}(n) = U_{n ,  1} \, U_{n+\hat{1} ,  2} \,
U_{n+\hat{2},1}^{\dagger} \, U_{n, 2}^\dagger$
with $\hat{\mu}$ being the unit vector 
in the $\mu$ direction $(\mu=1,2)$.
This system can be solved analytically
using the character 
expansion \cite{Balian:1974ts,Migdal:1975zg,Rothe:1992nt}, 
and the partition function is given 
by 
\begin{align}
  Z = \sum_{n=1}^\infty \left[ \frac{2}{\beta} I_n(\beta)\right]^V \ ,
  \label{Z 2dYM}
\end{align}
where $I_n(x)$ is the modified Bessel function of the first kind 
and $V=N_{\rm t} N_{\rm s}$ is the number of sites on the lattice.

The parameter $\beta$ in (\ref{eq:action_orig})
is related to the gauge coupling constant $g$ by $\beta= 1/g^2$,
and it is usually taken to be real positive,
in which case the action is real and the sign problem does not occur.
Note, however, that the model itself is well defined also 
for complex $\beta$, in which case
the action is complex and the sign problem occurs.
Since the analytic solution (\ref{Z 2dYM}) remains valid
for complex $\beta$, this simple gauge theory serves as a 
useful testing ground for methods which aim at solving the sign problem.

In order to apply the CLM to the 2dYM with complex $\beta$,
we complexify the link variables as 
$U_{n \mu} \mapsto {\cal U}_{n \mu} \in  {\rm SL}(N_{\rm c},\bbC)$
and extend the action and the observables to 
holomorphic functions of the complexified variables. 
For instance, the plaquette is extended to 
\begin{align}
  U_{12}(n) \mapsto {\cal U}_{12}(n)
  = {\cal U}_{n,1} \, {\cal U}_{n+\hat{1},2}\, 
{\cal U}_{n+\hat{2},1}^{-1}\, {\cal U}_{n,2}^{-1} \ ,
\end{align}
and the action is extended to
\begin{align}
S(U) \mapsto S({\cal U}) &=-\frac{\beta}{2N_{\rm c}}
\sum_n {\rm tr} \biggl[ {\cal U}_{12}(n) + {\cal U}_{12}^{-1}(n)\biggr] \ .
\label{eq:comp_action}
\end{align}
Note that the Hermitian conjugate of $U_{n\mu}$ is replaced with 
the inverse of ${\cal U}_{n\mu}$ so that the action becomes
a holomorphic function of the complexified link variables ${\cal U}_{n\mu}$.

A fictitious time evolution of
the complexified link variables is
defined by the complex Langevin equation 
with gauge cooling as
\begin{alignat}{3}
\widetilde{\cal U}_{n \mu} (t) & = 
g_{n} \, 
{\cal U}_{n \mu} (t) \, 
g_{n+\hat{\mu}}^{-1}   \ ,  
\label{eq:Langevin-discretized2-complexified-cooled0-lgt}
\\
{\cal U}_{n \mu} (t+\epsilon) & = 
\exp \Big\{
i \sum_a 
\Big( \epsilon v_{a n \mu} (\widetilde{\cal U}(t)) 
+ \sqrt{\epsilon} \eta_{a n \mu}(t) \Big)
\, t_a \Big\}
\,  \widetilde{\cal U}_{n \mu} (t)  \ ,
\label{eq:Langevin-discretized2-complexified-cooled-lgt}
\end{alignat}
where $t$ is the discretized Langevin time
with a stepsize $\epsilon$.
The SU($N_{\rm c}$) generators $t_a \ (a=1,\cdots, N_{\rm c}^2-1)$ are 
normalized as $\tr (t_a t_b)=\delta_{ab}$
and the real Gaussian noise $\eta_{a n \mu}(t)$ is normalized 
as $\langle \eta_{a' n' \mu'}(t') \eta_{ a n \mu} (t) 
\rangle_\eta = 2 \delta_{a' a} \delta_{n'n} \delta_{\mu'\mu} \delta_{t' t}$. 
The drift term $v_{an\mu}$ is defined by 
\begin{align}
  v_{an\mu}(\mathcal U) = - {\cal D}_{an\mu} S({\cal U})
\equiv -\left. \frac{\partial}{\partial z}
  S(e^{i z t_a}\mathcal{U})\right|_{z=0} \ .
  \label{v 2dYM}
\end{align}

The gauge transformation 
in (\ref{eq:Langevin-discretized2-complexified-cooled0-lgt})
represents the gauge cooling, where $g_n$ takes values in ${\rm SL}(N_{\rm c},\bbC)$.
In this model, 
the complexified link variables can have large components
in the non-compact direction of SL(${N_{\rm c}}, \mathbb C$) \cite{Makino:2015ooa},
which represents the excursion problem.
We try to avoid this problem as much as possible
by using the gauge cooling.
For that purpose, we 
define
the unitarity norm, 
\begin{align}
{\cal N} = \frac{1}{2 V} \sum_{n,\mu} 
\mathrm{Tr}\left[ {\cal U}_{n\mu} {\cal U}_{n\mu}^\dagger 
  + {\cal U}_{n\mu}^{-1} ({\cal U}_{n\mu}^{-1}) ^\dagger \right] \ ,
\label{n}
\end{align}
which measures the deviation of the link variables from SU($N_{\rm c}$), 
and choose the gauge transformation
in (\ref{eq:Langevin-discretized2-complexified-cooled0-lgt})
in such a way that the norm ${\cal N}$ is minimized \cite{Seiler:2012wz}. 

Let us define the magnitude $u_n(\mathcal U)$ of the drift term 
at site $n$ by
\begin{align}
  u_n(\mathcal U) = \sqrt{\frac{1}{2(N_{\rm c}^2-1)}  
\sum_{\mu=1,2}\sum_{a=1}^{N_{\rm c}^2-1} |v_{an\mu}(\mathcal U)|^2}
\end{align}
with $v_{an\mu}(\mathcal U)$ being the drift term \eqref{v 2dYM}.
Then, the probability distribution of the drift term 
can be defined as
\begin{align}
p(u) = 
\int {\cal D U}  \, 
\frac{1}{V}\sum_n \delta ( u - u_n({\cal U}) ) P(\mathcal U) \ ,
\label{u-prob 2dYM}
\end{align}
where $P(\mathcal U)$
represents the probability distribution of $\mathcal U_{n\mu}(t)$
in the $t\rightarrow \infty$ limit.
This definition of $p(u)$ respects the ${\rm SU}(N_{\rm c})$ gauge symmetry
of the original theory.





\begin{figure}[htbp] 
\begin{center}
\includegraphics[width=7cm]{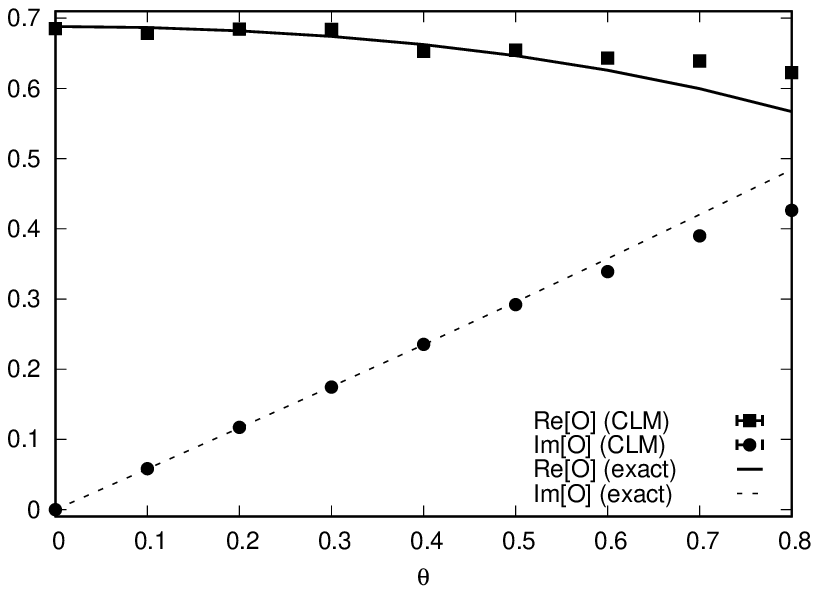}
\includegraphics[width=7cm]{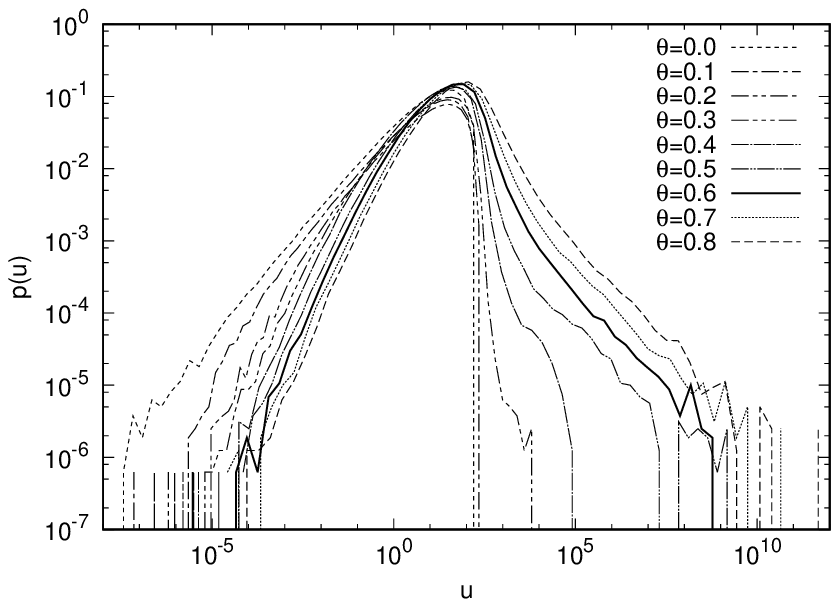}
\caption{(Left) The real and imaginary parts of 
the expectation value of the average plaquette 
obtained by the CLM
are plotted against $\theta$. The solid and dotted lines represent 
the exact results obtained by the character expansion.
(Right)
The probability distribution \eqref{u-prob 2dYM}
of the drift term 
is shown in log-log plots for various $\theta$.
}
\label{Fig:ave_plaq}
\end{center}
\end{figure}


\subsection{results}

Here we present our results obtained by the CLM.
Following \cite{Makino:2015ooa},
we choose $N_{\rm c}=2$, the lattice size $N_{\rm s} = N_{\rm t} = 4$
and $\beta = 1.5 \, e^{i \theta}$ with various $\theta$. 
The simulation was performed 
for the total Langevin time $t\sim 500$
with a fixed stepsize $\epsilon = 10^{-5}$.

In Fig.~\ref{Fig:ave_plaq} (Left), 
we show the expectation value of the average plaquette 
defined by 
\begin{align}
  {\cal O} = \frac{1}{V} \sum_n {\rm tr} \ {\cal U}_{12}(n) 
\end{align}
as a function of $\theta$, which agrees with the result
obtained in ref.~\cite{Makino:2015ooa}.
In particular, the CLM fails to reproduce the exact results 
for $\theta\gtrsim 0.6$.
As was reported in ref.~\cite{Makino:2015ooa},
the unitarity norm (\ref{n})
is under control for all the parameters investigated
and, in particular,
it does not blow up even for the cases in which the CLM gives wrong results.
In ref.~\cite{Makino:2015ooa}, the scatter plot of the average plaquette
was also studied, but the results for 
$\theta \gtrsim 0.6$
were not able to detect the existence of the excursion problem.

In Fig.~\ref{Fig:ave_plaq} (Right),
we show the probability distribution \eqref{u-prob 2dYM} 
of the drift term for various $\theta$.
We find that the distribution falls off exponentially or faster
for $\theta\lesssim 0.4$,
while it falls off only by a power law for $\theta\gtrsim 0.6$.
This implies that our criterion based on the
probability distribution of the drift term
can indeed tell the parameter region in which the CLM gives correct results.



\section{Chiral Random Matrix Theory for finite density QCD}
\label{sec:cRMT}

In this section, we consider the cRMT \cite{Osborn:2004rf,Bloch:2012bh},
which was proposed as a toy model for finite density QCD.
This model was studied by the CLM in 
refs.~\cite{Mollgaard:2013qra,Mollgaard:2014mga,Nagata:2015ijn,Nagata:2016alq},
and it was found that the singular-drift problem 
occurs in some parameter region.

\subsection{the model}

The partition function of the model is given by \cite{Bloch:2012bh}
\begin{align}
Z &= \int d\Phi_1d\Phi_2 \,  [\det (D+m)]^{N_{\rm f}} e^{-S_{\rm b}}  \ ,
\label{crmt}
\end{align}
where $N_{\rm f}$ is the number of flavors and $m>0$ represents
the degenerate quark mass.
%
The dynamical variables consist of two general $N\times (N+\nu)$ complex matrices
$\Phi_{k} \ (k=1,2)$,
where the integer $\nu$ represents the topological index.
The action $S_{\rm b}$ in (\ref{crmt}) is given by 
\begin{align}
S_{\rm b}&=2N \sum_{k=1}^{2}{\rm Tr} (\Psi_k \Phi_k) \ ,
\label{Sb} 
\end{align}
where $\Psi_k \ (k=1,2)$ are $(N+\nu)\times N$ matrices defined by
\begin{align}
\Psi_k=(\Phi_k)^\dagger  \ .
\label{hc-constraint}
\end{align}
The reason for introducing new matrices representing
the Hermitian conjugate of $\Phi_k$ will be clear shortly.
The Dirac operator $D$ in (\ref{crmt}) is given by 
\begin{align}
D&=\left( \begin{matrix}
0 & X \\
Y & 0
\end{matrix} \right)  \ , \quad \quad \quad
\left\{
\begin{array}{ccc}
X & = & e^{\mu} \Phi_1 + e^{-\mu} \Phi_2  \\
Y & = & -e^{-\mu} \Psi_1 - e^\mu \Psi_2  \ ,
\end{array}
\right.
\label{Eq:2015Sep01eq2}
\end{align}
where $\mu$ is the chemical potential.
The effective action of this model reads
\begin{align}
S_{\rm eff}&= S_{\rm b} - N_{\rm f} \ln \det (D+m) \ .
\label{Seff} 
\end{align}

When one tries to apply Monte Carlo methods
to this model, the sign problem occurs for $\mu \neq 0$ due to the
complex fermion determinant $\det (D+m)$.
To see that, let us note first that
the Dirac operator $D$ satisfies the relation
\begin{align}
D \gamma_5  & = - \gamma_5 D  \ , \quad \quad \quad
 \gamma_5 =\begin{pmatrix} {\bf 1}_{N} & 0 \\ 0 & -{\bf 1}_{N+\nu} \end{pmatrix}
\label{anticommuting}
\end{align}
for any $\mu$.
This implies that all the nonzero eigenvalues of $D$ are paired 
with the ones with the sign flipped.
When $\mu=0$, $D$ is anti-Hermitian
and its eigenvalues are purely imaginary,
which implies that
the determinant $\mathrm{det}(D+m)$ is real semi-positive.
On the other hand, when $\mu\neq 0$, $D$ is no longer anti-Hermitian
and its eigenvalues can take complex values.
In this case, the determinant $\mathrm{det}(D+m)$ 
is complex in general, which causes the sign problem.
Since the model is actually analytically solvable, it serves 
as a useful toy model for investigating the sign problem
that occurs in finite density QCD.

Let us apply the CLM to the cRMT with $\mu\neq 0$.
First we consider real variables corresponding to
the real part and the imaginary part of $(\Phi_k)_{ij}$ 
and complexify these variables.
The action and the observables are extended to 
holomorphic functions of these complexified variables
by analytic continuation.
It is easy to convince oneself that
this simply amounts to disregarding the constraint (\ref{hc-constraint})
and extending the action and the observables to
holomorphic functions of $\Phi_k$ and $\Psi_k$ ($k=1,2$).

A fictitious time evolution of the complex matrices
$\Phi_k$ and $\Psi_k$ ($k=1,2$) is given by 
the complex Langevin equation with gauge cooling as
\begin{align}
  \tilde \Phi_k(t) &= g \, \Phi_k(t) \, h^{-1} \ ,
\quad \tilde \Psi_k(t)=h \, \Psi_k(t) \, g^{-1} \ ,
\label{cle-cRMT-gc}
\\
  \Phi_{k}(t+\epsilon)
  &= \epsilon\left[-2N\tilde \Phi_{k}(t)
  - N_{\rm f} e^{(-1)^{k}\mu} \, W^{-1}(\tilde\Phi(t),\tilde\Psi(t)) \, 
   X(\tilde\Phi(t))\right]
  + \sqrt{\epsilon} \, \eta_{k}(t) \ , \nonumber\\
  \Psi_{k}(t+\epsilon)
  &= \epsilon\left[-2N\tilde \Psi_{k}(t)
  + N_{\rm f} e^{(-1)^{k+1}\mu} \, 
  Y(\tilde\Psi(t)) \, W^{-1}(\tilde\Phi(t),\tilde\Psi(t)) \right]
  + \sqrt{\epsilon}\, \eta_{k}^\dagger(t) \ ,
\label{cle-cRMT}
\end{align}
where $W=m^2-XY$ is an $N\times N$ matrix.
The $N\times (N+\nu)$ matrices $\eta_{k}(t)$
have components taken from complex Gaussian variables
normalized by
$\langle \eta_{k,ij}(t) \eta_{k',i'j'}^{*} (t') \rangle_\eta
= 2 \delta_{kk'} \delta_{ii'} \delta_{jj'} \delta_{tt'}$. 
Eq.~(\ref{cle-cRMT-gc})
represents the gauge cooling with 
$g\in \mathrm{GL}(N,\mathbb{C})$ and $h\in \mathrm{GL}(N+\nu,\mathbb{C})$,
which are obtained by complexifying the 
$\mathrm{U}(N)\times \mathrm{U}(N+\nu)$ symmetry of the original 
model (\ref{crmt}).

In ref.~\cite{Mollgaard:2013qra}, the same model was studied
by the CLM without gauge cooling, and 
it was found that one obtains wrong results 
for small quark mass or large chemical potential.
The reason for this failure is the singular-drift 
problem \cite{Nishimura:2015pba}, which occurs
due to eigenvalues of $(D+m)$ close to zero.
In ref.~\cite{Nagata:2016alq},
we proposed to use the gauge cooling 
to avoid the singular drift problem as much as possible.\footnote{While
the gauge cooling is found to be useful in avoiding the singular drift problem
in the present model, it is found to be ineffective
in a similar model \cite{Stephanov:1996ki} according to 
a recent study \cite{Bloch:2017sex}.}
There, three different types of ``norm'' 
were considered
so that the gauge transformations $g$ and $h$ in (\ref{cle-cRMT-gc})
can be determined by minimizing them.
A counterpart of the unitarity norm (\ref{n}) in gauge theory,
which is called the Hermiticity norm,
can be defined as
\begin{align}
\mathcal N_{\rm H} &= \frac{1}{N}\, \sum_{k=1,2}  {\rm Tr}
[ (\Psi_k-\Phi_k^\dagger)^\dagger (\Psi_k-\Phi_k^\dagger) ] \ ,  
\end{align}
which measures the violation of the relation (\ref{hc-constraint}).
It turned out, however, that the gauge cooling with this norm
does not reduce the singular drift problem.
This led us to consider
a norm that 
is related directly to the eigenvalue distribution of the Dirac operator.
A simple choice is given by
\begin{align}
\mathcal N_1 = \frac{1}{N}{\rm Tr}
\left[(X+Y^\dagger)(X+Y^\dagger)^\dagger\right] \ ,
\label{normtype1}
\end{align}
which measures the deviation of the Dirac operator 
(\ref{Eq:2015Sep01eq2})
from an anti-Hermitian matrix.
The gauge cooling with this norm has an effect of 
making the eigenvalue distribution 
of $D+m$ narrower in the real direction.
Another choice is given by
\begin{align}
\mathcal N_{2} = \sum_{a=1}^{n_{\rm ev}} e^{-\xi \alpha_a} \ ,
\label{normtype2}
\end{align}
where $\xi$ is a real parameter
and $\alpha_a$ are the real semi-positive eigenvalues of $M^\dagger M$ 
with $M=D+m$.
In eq.~(\ref{normtype2}), we take a sum over the $n_{\rm ev}$ 
smallest eigenvalues of $M^\dagger M$.
The gauge cooling with this norm has an effect of 
achieving $\alpha_a \gtrsim 1/\xi$ and suppressing the
appearance of small $\alpha_a$.
Since $\alpha_a \gtrsim 1/\xi$ implies $|\lambda_a|^2 \gtrsim 1/\xi$, 
where $\lambda_a$ are the eigenvalues of $M$,
it is also expected to suppress the appearance of $\lambda_a$ close to zero.
In some cases, 
the use of the norm (\ref{normtype1}) or (\ref{normtype2}) 
causes
the excursion problem.
In order to avoid this, 
we consider a combined norm
\begin{align}
\hat{\mathcal N}_{i}(s) = s \mathcal N_{\rm H} + (1-s) \mathcal N_i 
\quad \quad \quad
\mbox{for~}i=1,2 \ ,
\label{Ntot}
\end{align}
where
$s$ $(0 \le s \le 1)$ is a tunable parameter.


Let us discuss how we define the probability distribution
of the drift term.
Here we set the topological index $\nu=0$ for simplicity
so that the dynamical variables $\Phi_k$ and $\Psi_k$ are 
$N \times N$ square matrices.
We denote the drift terms of $\Phi_k$ and $\Psi_k$ by $F_k$ 
and $G_k$ ($k=1,2$), and represent
the eigenvalues of  $(F_k^\dagger F_k)^{1/2}$ and 
$(G_k^\dagger G_k)^{1/2}$ by 
$v_k^{(a)}$ and $w_k^{(a)}$ ($a=1, \cdots , N$), respectively.
Then we define the probability distribution 
of the drift term as
\begin{align}
p(u)=\frac{1}{2N}\int \prod_{k=1,2}d\Phi_k d\Psi_k
\sum_{k=1}^{2}\sum_{a=1}^{N}
\left(\delta(u-v^{(a)}_k(\Phi,\Psi))
+\delta(u- w^{(a)}_k(\Phi,\Psi))\right)  P(\Phi,\Psi) \ ,
\label{p of crmt}
\end{align}
where $P(\Phi,\Psi)$
is the probability distribution of $\Phi_k(t)$ and $\Psi_k(t)$
in the $t\rightarrow \infty$ limit.
This definition of $p(u)$
respects the $U(N)\times U(N)$ symmetry of the original theory.



\begin{figure}[htbp]
\centering
\includegraphics[width=7cm]{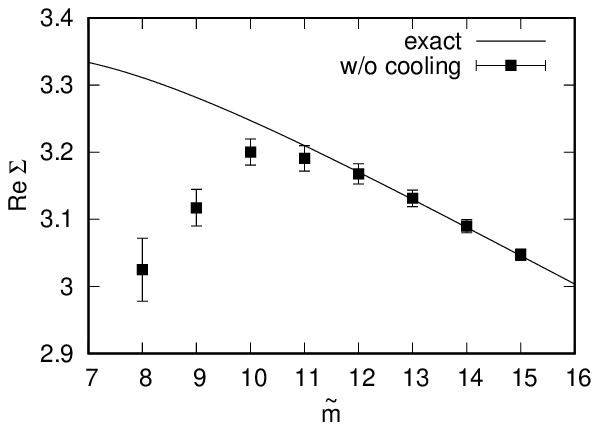}
\includegraphics[width=7cm]{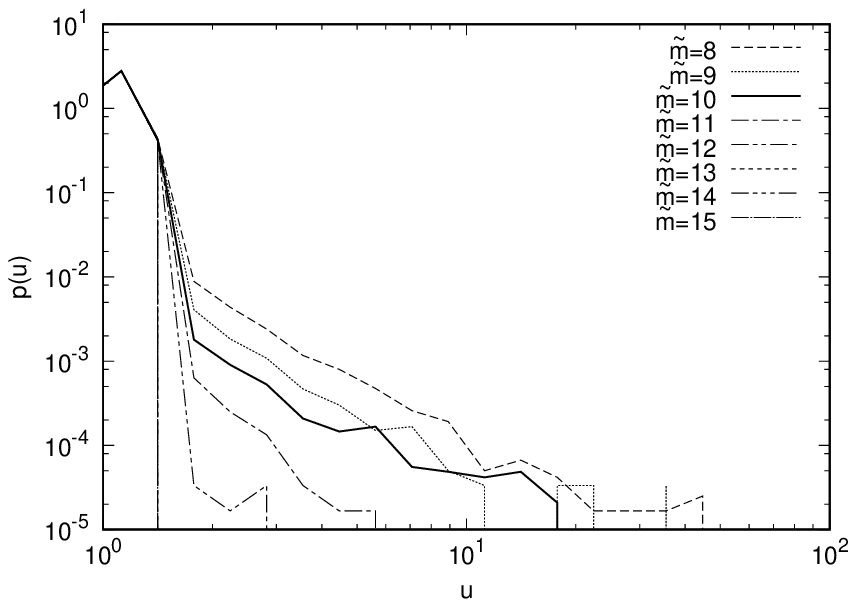}\\
\includegraphics[width=7cm]{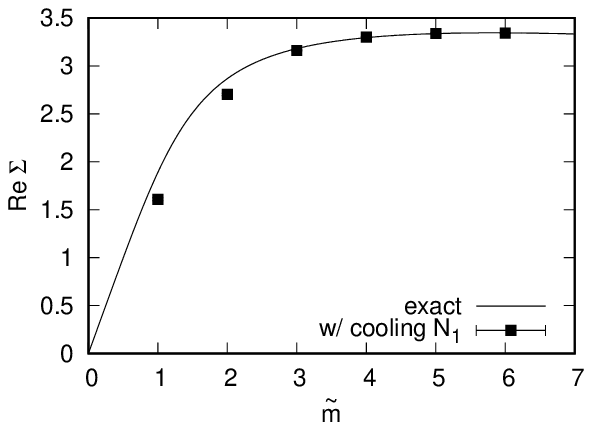}
\includegraphics[width=7cm]{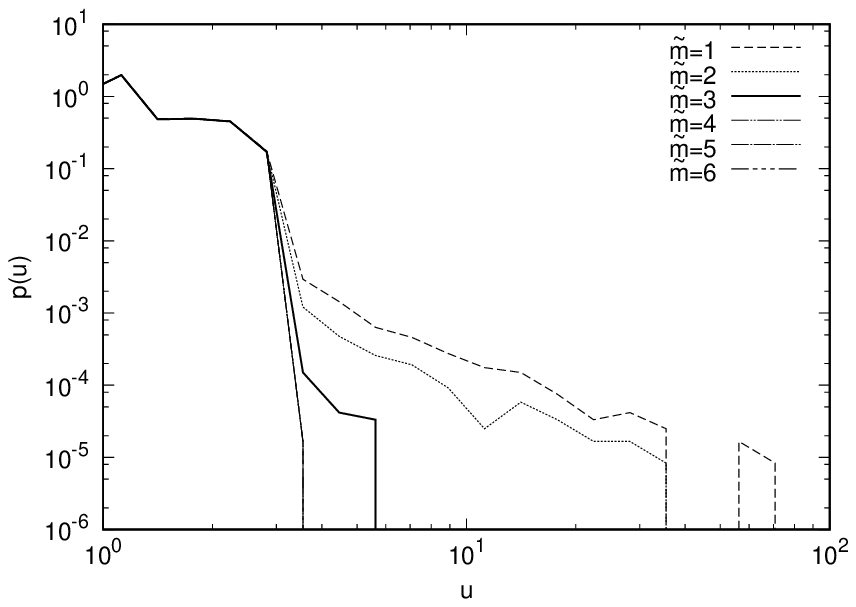}\\
\includegraphics[width=7cm]{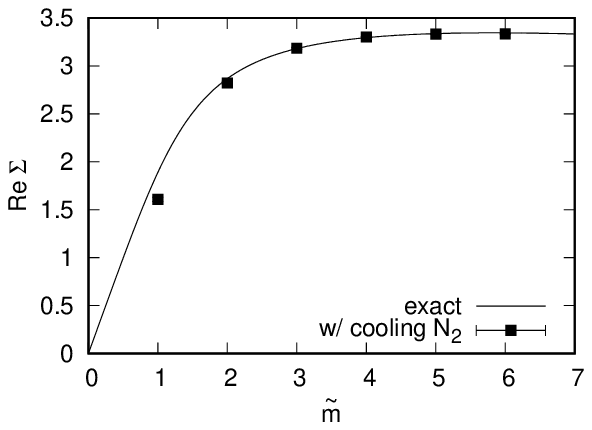}
\includegraphics[width=7cm]{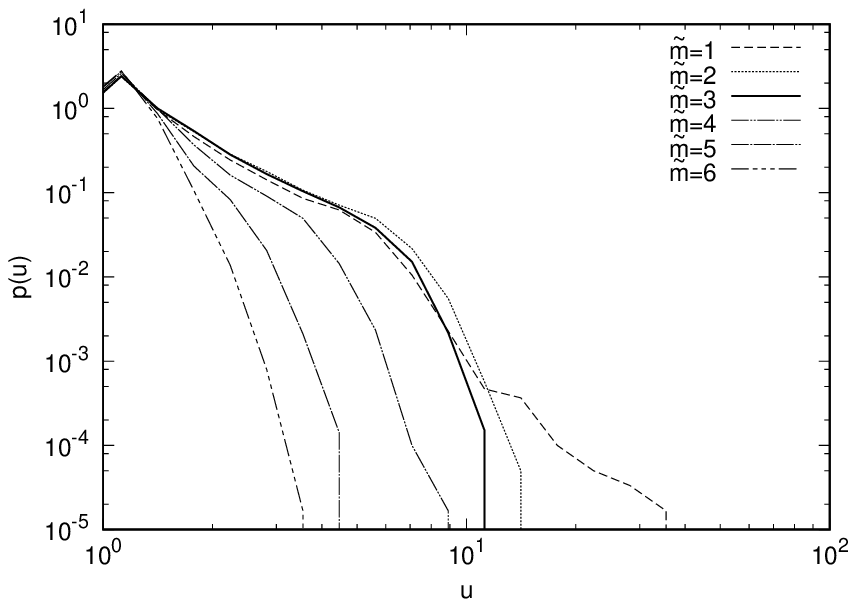}
\caption{
  (Left) The real part of the chiral condensate is plotted against
  $\tilde m$
  for the cases without gauge cooling (Top) and
  with gauge cooling using the norms $\hat{\mathcal N}_1$ (Middle)
  and $\hat{\mathcal N}_2$ (Bottom).
 (Right) The probability distribution (\ref{p of crmt}) of 
 the drift term is
  shown in log-log plots for each case.
}
\label{crmt fig2}
\end{figure}

\subsection{results}

Here we present our results obtained by the CLM.
Following previous works \cite{Mollgaard:2014mga,Nagata:2016alq},
we choose
$\nu=0$, $N_{\rm f}=2$, $N=30$, $\tilde \mu\equiv \mu \sqrt{N}=2$ 
with various $\tilde m \equiv mN$.
The simulation was performed 
for the total Langevin time $t=10$
with a fixed stepsize $\epsilon = 5\times 10^{-5}$.
For gauge cooling, we use the combined norm 
$\hat{\mathcal N}_i(s)$ defined in \eqref{Ntot} with $s=0.01$.
%

In Fig.~\ref{crmt fig2} (Left), we show the real part of the chiral condensate
\begin{align}
\Sigma = \frac{1}{N} \frac{\partial}{\partial m} \log Z 
\label{chiral}
\end{align}
obtained by the CLM with or without gauge cooling and compare it
with the exact result given in ref.~\cite{Osborn:2004rf}.
In the case without gauge cooling (Top),
we find that the result of the CLM deviates from the 
exact result for $\tilde m \lesssim 11$.
On the other hand, in the case with gauge cooling
using the norm $\hat{\mathcal N}_1$ (Middle),
we find that the result of the CLM deviates from the 
exact result for $\tilde m \lesssim 2$.
When we use the norm $\hat{\mathcal N}_2$ (Bottom) instead,
the deviation occurs for $\tilde m\lesssim 1$.

In Fig.~\ref{crmt fig2} (Right),
we show the probability distribution $p(u)$ of the drift term
obtained by the CLM with or without gauge cooling
focusing on the parameter region in which the result starts to deviate
from the exact result.
In the case without gauge cooling (Top),
we find 
that 
the probability distribution falls off exponentially 
or faster for $\tilde m\gtrsim 12$,
while it develops a power-law tail for $\tilde m \lesssim 11$.
%
In the case with gauge cooling using the norm $\hat{\mathcal N}_1$ (Middle),
we find that the probability distribution falls off exponentially 
or faster for $\tilde m\gtrsim 3$,
while it develops a power-law tail for $\tilde m\lesssim 2$.
When we use the norm $\hat{\mathcal N}_2$ (Bottom) instead,
the probability distribution falls off exponentially or faster 
for $\tilde m\gtrsim 2$,
while it develops a power-law tail for $\tilde m\lesssim 1$.
This implies that the probability distribution of the drift term
can indeed tell the parameter region in which the CLM gives correct results.


\section{Summary and discussions}
\label{sec:summary}

In this paper we have shown
that the probability distribution of the drift term 
indeed provides a useful criterion for judging the reliability 
of results obtained by the CLM.
According to our criterion, the CLM gives correct results 
when the probability distribution of the drift term
falls off exponentially or faster.
We have tested it in two solvable
models with many dynamical degrees of freedom, i.e.,
2dYM with a complex coupling and the cRMT for finite density QCD.
While the CLM was known to fail in these models in some parameter region
due to the excursion problem and the singular-drift problem, respectively,
it was not possible to tell precisely
in which region the CLM fails without knowing the exact results.
Our criterion was able to determine this parameter region clearly.
Note also that the two apparently different problems can be
detected by a single criterion in a unified manner.

While it is widely appreciated that the CLM enables explicit 
calculations in various interesting models with the sign problem
at least in certain parameter
regions, its usefulness would be rather limited if there is no way
to tell precisely in which parameter region it works.
The establishment of such a criterion that enables this is 
therefore of particular importance.
It should be also emphasized that
our criterion requires essentially no additional cost
since the drift term is calculated anyway at each step 
in the Langevin simulation.
Indeed our criterion has played a crucial role in investigating
the spontaneous symmetry breaking in matrix 
models \cite{Ito:2016efb,Anagnostopoulos:2017gos}
motivated by superstring theory.
We consider that our criterion is indispensable 
in applying the CLM to many other interesting systems such as 
finite density QCD \cite{Nagata:2017pgc}.

\acknowledgments
We thank J.~Bloch, Y.~Ito, K.~Moritake, S.~Tsutsui and J.J.M.~Verbaarschot
for valuable discussions.
K.~N.\ and J.~N.\ were supported in part 
by Grant-in-Aid for Scientific Research (No.~26800154 and 16H03988, 
respectively)
from Japan Society for the Promotion of Science. 
S.~S.\ was supported by the MEXT-Supported Program 
for the Strategic Research Foundation at Private Universities 
``Topological Science'' (Grant No.~S1511006).

\bibliographystyle{JHEP}
\bibliography{ref_cle-2dym2}

\end{document}